\documentclass[aps,pof,preprint,superscriptaddress]{revtex4-1}

\usepackage{amsmath}

\usepackage[latin1]{inputenc}

\usepackage{color}
\usepackage{cancel}
\usepackage{hyperref}

\begin{document} 

\title{Exact equations for structure functions and equations for source terms up to the sixth order}

% repeat the \author .. \affiliation  etc. as needed
% \email, \thanks, \homepage, \altaffiliation all apply to the current
% author. Explanatory text should go in the []'s, actual e-mail
% address or url should go in the {}'s for \email and \homepage.
% Please use the appropriate macro foreach each type of information

% \affiliation command applies to all authors since the last
% \affiliation command. The \affiliation command should follow the
% other information
% \affiliation can be followed by \email, \homepage, \thanks as well.
\author{Norbert Peters}
\email[]{n.peters@itv.rwth-aachen.de}
\affiliation{Institut f\"ur Technische Verbrennung, RWTH Aachen University, Germany}

\author{Jonas Boschung}
%\email[]{j.boschung@itv.rwth-aachen.de}
\affiliation{Institut f\"ur Technische Verbrennung, RWTH Aachen University, Germany}

\author{Michael Gauding}
%\email[]{}
\affiliation{Chair of Numerical Thermo-Fluid Dynamics, TU Bergakademie Freiberg, Germany}

\author{Jens Henrik G\"obbert}
%\email[]{}
\affiliation{J\"ulich Aachen Research Alliance (JARA-HPC), RWTH Aachen University, Germany}

\author{Heinz Pitsch}
%\email[]{}
\affiliation{Institut f\"ur Technische Verbrennung, RWTH Aachen University, Germany}

\date{\today}

\begin{abstract}
We derive equations for the source terms appearing in structure function equations for the fourth and sixth order under the assumption of homogeneity and isotropy. The source terms can be divided into two classes, namely those stemming from the viscous term and those from the pressure term in the structure function equations. Both kinds are unclosed.
\end{abstract}

\maketitle

\section{Introduction}

Since Kolmogorov derived asymptotic results for the second order structure function $\left< \Delta u^2 \right>$ in the dissipation range and for the third order structure function $\left< \Delta u^3 \right>$ in the inertial range in 1941 from the Navier-Stokes equations under the assumption of homogeneity, incompressibility and isotropy, a vast amount of work was carried out on the subject. One of the reasons is that the scale of with the viscosity $\nu$ and the mean dissipation $\left< \varepsilon \right>$ in these results was thought to be universal for all kind of flows. Soon, it has been found that higher orders do not follow the scaling predicted by Kolmogorov and the deviations from it become stronger for higher orders. Hill~\cite{hill2001equations} and Yakhot~\cite{yakhot2001mean} independently derived higher order struture function equations. However, the source terms are unclosed. For more insight, we derive equations for the source terms. Applications will be presented in papers to be submitted and the present article is meant as a basis for these.

\section{Fourth order structure function equations}

We start be deriving the equations for higher order structure functions, the derivation of which has already been carried out by Hill\cite{hill2002exact,hill2001mathematics}. The reason for re-deriving them here is twofold. Firstly, the present article serves as basis for papers to appear in the refereed journals and for that purpose it is reasonable to have all equations in one document. Secondly, additional steps used in the derivation of the structure function equations are required for the derivation of their source term equations.

Only the steady state form of the averaged equations will be considered. The two points are denoted by $\boldsymbol{x} = (x_1,x_2,x_3)$  and $\boldsymbol{x}' = (x_1',x_2',x_3')$. Assuming incompressible flow, the momentum equations in the Navier-Stokes equations are written at the two points
\begin{equation}
\begin{aligned}
\frac{\partial u_i}{\partial t} + u_n \frac{\partial u_i}{\partial x_n} = - \frac{\partial p}{\partial x_i} + \nu \frac{\partial^2 u_i}{\partial x_j^2} , \ \ \ i=1,2,3
\label{eq_1}
\end{aligned}
\end{equation}
\begin{equation}
\begin{aligned}
\frac{\partial u_i'}{\partial t} + u_n' \frac{\partial u_i'}{\partial x_n'} = - \frac{\partial p'}{\partial x_i'} + \nu \frac{\partial^2 u_i'}{\partial x_j'^2} , \ \ \ i=1,2,3
\label{eq_2}
\end{aligned}
\end{equation}
Here $u_i$ and $u_i'$ are the components of the velocity, $p$ is the pressure and $\nu$ the kinematic viscosity. Einstein's summation convention for indices appearing twice is used. The density was set equal to unity. These equations are completed by the continuity equation which holds at both points,
\begin{equation}
\begin{aligned}
\frac{\partial u_i}{\partial x_i} = 0
\end{aligned}
\end{equation}
\begin{equation}
\begin{aligned}
\frac{\partial u_i'}{\partial x_i'} = 0
\end{aligned}
\end{equation}
Substracting eq.~\eqref{eq_2} from eq.~\eqref{eq_1} one obtains for the velocity increment $\Delta u_i$ defined by $\Delta u_i = u_i(\boldsymbol{x}) - u_i'(\boldsymbol{x}')$ the equation 
\begin{equation}
\begin{aligned}
\frac{\partial \Delta u_i}{\partial t} + u_n \frac{\partial \Delta u_i}{\partial x_n} + u_n' \frac{\partial \Delta u_i}{\partial x_n'} = - \underbrace{ \left( \frac{\partial p}{\partial x_i} - \frac{\partial p'}{\partial x_i'} \right) }_{\Delta P_i} + \nu \left( \frac{\partial^2 \Delta u_i}{\partial x_n^2} + \frac{\partial^2 \Delta u_i}{\partial x_n'^2} \right)
\label{eq_6}
\end{aligned}
\end{equation}
Here the difference of the pressure gradient at the two points is defined as $\Delta P_i$ and the pressure is given by the Poisson equation
\begin{equation}
\begin{aligned}
\frac{\partial^2 p}{\partial x_n^2} + \frac{\partial u_i}{\partial x_j} \frac{\partial u_{j}}{\partial x_i} = 0
\end{aligned}
\end{equation}
From this an equation for the pressure gradient $\partial p / \partial x_k$ may be derived by diferentiation
\begin{equation}
\begin{aligned}
\frac{\partial^2}{\partial x_n^2} \left( \frac{\partial p}{\partial x_i} \right) + \underbrace{ \left( \frac{\partial^2 u_i}{\partial x_j \partial x_k} \frac{\partial u_j}{\partial x_i} + \frac{\partial u_i}{\partial x_j} \frac{\partial^2 u_j}{\partial x_i \partial x_k} \right) }_{R_k} = 0
\label{eq_8}
\end{aligned}
\end{equation}
which defines the operator $R_k$. The equation for $\Delta P_k$, multiplied by the viscosity, then becomes
\begin{equation}
\begin{aligned}
\nu \left(  \frac{\partial^2 \Delta P_k}{\partial x_n^2} + \frac{\partial^2 \Delta P_k}{\partial x_n'^2} \right) + \nu \left( R_k - R_k' \right) = 0
\end{aligned}
\end{equation}
Equation~\eqref{eq_6} is then multiplied by $\Delta u_j$ and the corresponding equation for $\Delta u_j$ is multiplied by $\Delta u_i$ to obtain
\begin{equation}
\begin{aligned}
\Delta u_j \frac{\partial \Delta u_i}{\partial t} + u_n \Delta u_j \frac{\partial \Delta u_i}{\partial x_n} + u_n' \Delta u_j \frac{\partial \Delta u_i}{\partial x_n'} = - \Delta u_j \Delta P_i + \nu \Delta u_j \left( \frac{\partial^2 \Delta u_i}{\partial x_n^2} + \frac{\partial^2 \Delta u_i}{\partial x_n'^2} \right)
\label{eq_10}
\end{aligned}
\end{equation}
\begin{equation}
\begin{aligned}
\Delta u_i \frac{\partial \Delta u_j}{\partial t} + u_n \Delta u_i \frac{\partial \Delta u_j}{\partial x_n} + u_n' \Delta u_i \frac{\partial \Delta u_j}{\partial x_n'} = - \Delta u_i \Delta P_j + \nu \Delta u_i \left( \frac{\partial^2 \Delta u_j}{\partial x_n^2} + \frac{\partial^2 \Delta u_j}{\partial x_n'^2} \right)
\label{eq_11}
\end{aligned}
\end{equation}
Adding eq.~\eqref{eq_10} and eq.~\eqref{eq_11} one obtains an equation for $\Delta u_i \Delta u_j$,
\begin{equation}
\begin{aligned}
\frac{\partial \Delta u_i \Delta u_j}{\partial t} + u_n \frac{\partial \Delta u_i \Delta u_j}{\partial x_n} + u_n' \frac{\partial \Delta u_i \Delta u_j}{\partial x_n'} = &- \Delta u_j \Delta P_i - \Delta u_i \Delta P_j \\
&+ \nu \left( \frac{\partial^2 \Delta u_i \Delta u_j}{\partial x_n^2} + \frac{\partial^2 \Delta u_i \Delta u_j}{\partial x_n'^2} \right) \\
&- \left( \varepsilon_{ij} + \varepsilon_{ij}' \right) ,
\label{eq_12}
\end{aligned}
\end{equation}
where we used
\begin{equation}
\begin{aligned}
2 \nu \left(\frac{\partial \Delta u_i}{\partial x_n} \frac{\partial \Delta u_j}{\partial x_n} + \frac{\partial \Delta u_i}{\partial x_n'} \frac{\partial \Delta u_j}{\partial x_n'} \right) =  2 \nu \frac{\partial u_i}{\partial x_n}\frac{\partial u_j}{\partial x_n} + \frac{\partial u_i'}{\partial x_n'}\frac{\partial u_j'}{\partial x_n'}  = \varepsilon_{ij} + \varepsilon_{ij}' ,
\end{aligned}
\end{equation}

This also defines the instantaneous values of the dissipation $\varepsilon_{ij}$ at $\boldsymbol{x}$ and $\varepsilon_{ij}'$ at $\boldsymbol{x}'$. Equation~\eqref{eq_12} is an equation for velocity differences at two points. However, the derivatives do not reflect this and are still carried out with respect to $\boldsymbol{x}$ and $\boldsymbol{x}'$. Changing the independent variables from $\boldsymbol{x}$ and $\boldsymbol{x}'$ to the new independent variables 
\begin{equation}
\begin{aligned}
\boldsymbol{X} = \frac{1}{2} \left( \boldsymbol{x} + \boldsymbol{x}' \right) , \ \ \ \boldsymbol{r} = \left( \boldsymbol{x} - \boldsymbol{x}' \right)
\end{aligned}
\end{equation}
and using the transformation rules
\begin{equation}
\begin{aligned}
\frac{\partial}{\partial x_i} = \frac{\partial}{\partial r_i} + \frac{1}{2} \frac{\partial}{\partial X_i} , \ \ \ \frac{\partial}{\partial x_i'} = - \frac{\partial}{\partial r_i} + \frac{1}{2} \frac{\partial}{\partial X_i}
\end{aligned}
\end{equation}
one obtains for the transport term in eq.~\eqref{eq_12}
\begin{equation}
\begin{aligned}
u_n \frac{\partial \Delta u_i \Delta u_j}{\partial x_n} + u_n' \frac{\partial \Delta u_i \Delta u_j}{\partial x_n'} = \frac{\partial \Delta u_n \Delta u_i \Delta u_j}{\partial r_n}
\end{aligned}
\end{equation}
where the derivatives $\partial / \partial X_n$ were neglected, as they vanish after averaging due to the assumption of homogeneity. Likewise the Laplacian in eq.~\eqref{eq_12}) becomes
\begin{equation}
\begin{aligned}
\frac{\partial^2 \Delta u_i \Delta u_j}{\partial x_n^2} + \frac{\partial^2 \Delta u_i \Delta u_j}{\partial x_n'^2} = 2 \frac{\partial^2 \Delta u_i \Delta u_j}{\partial r_n^2} .
\end{aligned}
\end{equation}
We now average eq.~\eqref{eq_12}. Defining the second order longitudinal structure functions as
\begin{equation}
\begin{aligned}
S_{2,0} = \left< \Delta u_1^2 \right>
\end{aligned}
\end{equation}
and the transverse structure function as
\begin{equation}
\begin{aligned}
S_{0,2} = \left< \Delta u_2^2 \right>
\end{aligned}
\end{equation}
and by setting $i=1$, $j=1$ and $i=2$, $j=2$, respectively, in eq.~\eqref{eq_12} one obtains the transport equations for $S_{2,0}$ and $S_{0,2}$. We can further simplify these equations under the assumption of isotropy. Specifically, we can find the isotropic form of both the gradient and the Laplacian, so that they depend on the separation distance $r$ only, instead of the separation vector $r_n$. For the steady state case after transforming the transport term and the Laplacian into their isotropic form (cf. Robertson~\cite{robertson1940invariant}), we have
\begin{align}
\left( \frac{\partial S_{3,0}}{\partial r} + \frac{2}{r} S_{3,0} \right) - \frac{4}{r} S_{1,2} &= 2 \nu \left[ \frac{\partial^2 S_{2,0}}{\partial r^2} + \frac{2}{r} S_{2,0} + \frac{4}{r^2} \left( S_{0,2} - S_{2,0} \right) \right] - 2 \left< \varepsilon_{11} \right> \label{eq_19}\\
\left( \frac{\partial S_{1,2}}{\partial r} + \frac{2}{r} S_{1,2} \right) &= 2 \nu \left[ \frac{\partial^2 S_{0,2}}{\partial r^2} + \frac{2}{r} S_{0,2} - \frac{2}{r^2} \left( S_{0,2} - S_{2,0} \right) \right] - 2 \left< \varepsilon_{22} \right> .
\label{eq_20}
\end{align}
These are the second order structure function equations given by Hill~\cite{hill2001mathematics}. In~\cite{hill2001mathematics} it is also noted that due to isotropy
\begin{equation}
\begin{aligned}
\left< \varepsilon_{11} \right> = \left< \varepsilon_{22} \right> = \frac{2}{3} \left< \varepsilon \right> .
\end{aligned}
\end{equation}
Continuity provides an additional relation between $S_{2,0}$ and $S_{0,2}$
\begin{equation}
\begin{aligned}
\frac{r}{2}\frac{\partial S_{2,0}}{\partial r} + S_{2,0} - S_{0,2} = 0
\end{aligned}
\end{equation}
as well as between $S_{3,0}$ and $S_{1,2}$
\begin{equation}
\begin{aligned}
r \frac{\partial S_{3,0}}{\partial r} + S_{3,0} - 6 S_{1,2} = 0
\end{aligned}
\end{equation}
Using these two equations in eq.~\eqref{eq_19}, this leads then to the well-known Kolmogorov equation~\cite{kolmogorov1941dissipation},
\begin{equation}
\begin{aligned}
\left( \frac{d}{dr} + \frac{4}{r} \right) S_{3,0} = 6 \nu \left( \frac{d}{dr} + \frac{4}{r} \right) \frac{d S_{2,0}}{dr} - 4 \left< \varepsilon \right> .
\end{aligned}
\end{equation}
Kolmogorov~\cite{kolmogorov1941dissipation} had derived two asymptotic solutions of this equation, namely $S_{2,0} = \left<\varepsilon \right> r^2 / (15 \nu)$ for $r \to 0$ and $S_{3,0} = 4/5 \left< \varepsilon \right> r$ in the inertial range. The latter is known as the four-fifth law. Next we consider equations for structure functions and equations for source terms at the fourth order. The procedure is the same as above, i.e. we multiply the equation for $\Delta u_i$ (eq.~\eqref{eq_6}) by $\Delta u_j \Delta u_k \Delta u_l$ and add up all four combinations resulting in an equation for $\Delta u_i \Delta u_j \Delta u_k \Delta u_l$. In the next step, the coordinate transform outlined above is used and the gradient and Laplacian are re-written in their isotropic form. This is presented in Appendix~A below in some detail. Hill\cite{hill2001equations} gave the necessary coefficients up to the seventh order and presented a matrix algorithm which can be used to derive the isotropic form for all orders. Note that the coefficients in the Laplacian of the higher order structure function equations given by Hill~\cite{hill2001mathematics} and consequently the matrix algorithm were corrected in \url{http://arxiv.org/abs/physics/0102055}. In their  isotropic form the equations then read
\begin{align}
\frac{\partial S_{5,0}}{\partial r} &+ \frac{2}{r} S_{5,0} - \frac{8}{r} S_{3,2} = -\left< T_{4,0} \right>  - \left< E_{4,0} \right> \nonumber \\
& + 2 \nu \left[ \frac{\partial^2 S_{4,0}}{\partial r^2} + \frac{2}{r} \frac{\partial S_{4,0}}{\partial r} - \frac{8}{r^2} S_{4,0} + \frac{24}{r^2} S_{2,2} \right] \\
\frac{\partial S_{3,2}}{\partial r} &+ \frac{4}{r} S_{3,2} - \frac{8}{3 r} S_{1,4} = -\left< T_{2,2} \right> - \left< E_{2,2} \right>\nonumber\\
&+ 2 \nu \left[ \frac{2}{r^2} S_{4,0} + \frac{\partial^2 S_{2,2}}{\partial r^2} + \frac{2}{r} \frac{\partial S_{2,2}}{\partial r} - \frac{14}{r^2} S_{2,2} + \frac{8}{3 r^2} S_{0,4} \right] \\
\frac{\partial S_{1,4}}{\partial r} &+ \frac{6}{r} S_{1,4} = -\left< T_{0,4} \right> - \left< E_{0,4} \right> \nonumber\\
& + 2 \nu \left[ \frac{12}{r^2} S_{2,2} + \frac{\partial^2 S_{0,4}}{\partial r^2} + \frac{2}{r} \frac{\partial S_{0,4}}{\partial r} - \frac{4}{r^2} S_{0,4}  \right ]
\end{align}
Note that we have three coupled equations for the three unknowns $S_{5,0}$, $S_{3,2}$ and $S_{1,4}$. Noticeably, this is the case for all \emph{even} orders, but not the odd orders. The instantaneous values of the source terms are given by
\begin{align}
T_{4,0} &= 4 \Delta u_1^3 \Delta P_1 \label{eq_28} \\
E_{4,0} &= 6 \Delta u_1^2 \left( \varepsilon_{11} + \varepsilon_{11}' \right) \label{eq_29} \\
T_{2,2} &= 2 \Delta u_2^2 \Delta u_1 \Delta P_1 + 2 \Delta u_1^2 \Delta u_2 \Delta P_2 \label{eq_30} \\
E_{2,2} &= \Delta u_2^2 \left( \varepsilon_{11} + \varepsilon_{11}' \right) + 4 \Delta u_1 \Delta u_2 \left( \varepsilon_{12} + \varepsilon_{12}' \right) + \Delta u_1^2 \left( \varepsilon_{22} + \varepsilon_{22}' \right) \label{eq_31} \\
T_{0,4} &= 4 \Delta u_2^3 \Delta P_2 \label{eq_32} \\
E_{0,4} &= 6 \Delta u_2^2 \left( \varepsilon_{22} + \varepsilon_{22}'  \right) . \label{eq_33}
\end{align}
Different to the second order equations, there are now pressure source terms $T_{i,j} \sim \Delta u^{i+j-1} \Delta P$ acting on the system of equations. They appear at all orders higher than the second, where they had dropped out due to isotropy (cf. Hill~\cite{hill2001mathematics}). Additionally, there are the dissipation source terms $E_{i,j}$. Note that again the second order is special, inasmuch as the averaged dissipation source terms equal the one-point quantities $\left< \varepsilon_{i,j} \right>$. For higher orders, this is not the case, i.e. $E_{i,j}$ is now clearly $r$-dependent. Consequently, it is not possible to find asymptotic solutions in the spirit of Kolmogorov without closure assumptions.

Therefore, it is worthwhile to study the higher order source terms in more detail, as anomalous scaling of the structure functions most likely originates from the source terms. For that reason, we proceed to derive the equations for all six source terms in the following.

\section{Fourth order pressure source terms}

Equations for the pressure source terms of the fourth order structure function equations, which have the generic forms $u_i u_j u_k \Delta P_l$, will be derived next. For that purpose we first derive an equation for the unaveraged product of three velocity increments by extending the procedure which led to eq.~\eqref{eq_12} to the product of three velocity increments. The result is
\begin{equation}
\begin{aligned}
\frac{\partial \Delta u_i \Delta u_j \Delta u_k}{\partial t} &+ u_n \frac{\partial \Delta u_i \Delta u_j \Delta u_k}{\partial x_n} + u_n' \frac{\partial \Delta u_i \Delta u_j \Delta u_k}{\partial x_n'} = \\
&- \Delta u_j \Delta u_k \Delta P_i - \Delta u_i \Delta u_k \Delta P_j - \Delta u_i \Delta u_j \Delta P_k \\
&+ \nu \left( \frac{\partial^2 \Delta u_i \Delta u_j \Delta u_k}{\partial x_n^2} + \frac{\partial^2 \Delta u_i \Delta u_j \Delta u_k}{\partial x_n'^2} \right) \\
&- \Delta u_i \left( \varepsilon_{jk} + \varepsilon_{jk}' \right) - \Delta u_j \left( \varepsilon_{ik} + \varepsilon_{ik}' \right) - \Delta u_k \left( \varepsilon_{ij} + \varepsilon_{ij}' \right)
\end{aligned}
\end{equation}
We multiply this equation by $\Delta P_l$ and convert the transport term as
\begin{equation}
\begin{aligned}
u_n \Delta P_l \frac{\partial \Delta u_i \Delta u_j \Delta u_k}{\partial x_n} + u_n' \Delta P_l \frac{\partial \Delta u_i \Delta u_j \Delta u_k}{\partial x_n'} &= u_n \frac{\partial \Delta u_i \Delta u_j \Delta u_k \Delta P_l}{\partial x_n} + u_n' \frac{\partial \Delta u_i \Delta u_j \Delta u_k \Delta P_l}{\partial x_n'} \\
&- \Delta u_i \Delta u_j \Delta u_k \left( u_n \frac{\partial \Delta P_l}{\partial x_n} + u_n' \frac{\partial \Delta P_l}{\partial x_n'} \right)
\end{aligned}
\end{equation}
where the second term is interpreted as a source term and is shifted to the right hand side of the equation. Adding eq.~\eqref{eq_8} multiplied by $\Delta u_i \Delta u_j \Delta u_k$ the equation then reads
\begin{align}
u_n \frac{\partial \Delta u_i \Delta u_j \Delta u_k \Delta P_l}{\partial x_n} &+ u_n' \frac{\partial \Delta u_i \Delta u_j \Delta u_k \Delta P_l}{\partial x_n'} = \nonumber \\
&- \Delta u_j \Delta u_k \Delta P_i \Delta P_l - \Delta u_i \Delta u_k \Delta P_j \Delta P_l - \Delta u_i \Delta u_j \Delta P_k \Delta P_l \nonumber \\
&+ \nu \left( \frac{\partial^2 \Delta u_i \Delta u_j \Delta u_k \Delta P_l}{\partial x_n^2} + \frac{\partial^2 \Delta u_i \Delta u_j \Delta u_k \Delta P_l}{\partial x_n'^2} \right) \nonumber \\
&+ \nu \left(R_l - R_l' \right) \Delta u_i \Delta u_j \Delta u_k \nonumber \\
&- \Delta P_l \frac{\partial \Delta u_i \Delta u_j \Delta u_k}{\partial t} \label{eq_36} \\
&+ \Delta u_i \Delta u_j \Delta u_k \left( u_n \frac{\partial \Delta P_l}{\partial x_n} + u_n' \frac{\partial \Delta P_l}{\partial x_n'} \right) \nonumber \\
& - 2 \nu \left( \frac{\partial \Delta u_i \Delta u_j \Delta u_k}{\partial x_n} \frac{\partial \Delta P_l}{\partial x_n} + \frac{\partial \Delta u_i \Delta u_j \Delta u_k}{\partial x_n'} \frac{\partial \Delta P_l}{\partial x_n'} \right) \nonumber \\
&- \Delta u_i \Delta P_l \left( \varepsilon_{jk} + \varepsilon_{jk}' \right) - \Delta u_j \Delta P_l \left( \varepsilon_{ik} + \varepsilon_{ik}' \right) - \Delta u_k \Delta P_l \left( \varepsilon_{ij} + \varepsilon_{ij}' \right) \nonumber
\end{align}
The unsteady  term has been retained and was shifted to the right hand side of this equation because it does not vanish in the steady state case after averaging. The transport terms and the diffusion terms in this equation are transformed into their isotropic form. For the transport term, we can not simply use the coefficients given by Hill, due to different symmetries of the tensor $\Delta u_k T_{i,j}$ compared to $S_{i,j}$. This derivation is somewhat laborious and reported in Appendix~B. Consequently, there are more scalar functions needed to determine the transport term. Specifically, two additional terms $\left< \Delta u_2 T_{3,1} \right>$ and $\left< \Delta u_2 T_{3,1} \right>$ appear. These are defined as
\begin{equation}
\left< \Delta u_2 T_{3,1} \right> = \left< \Delta u_2 \left( 3 \Delta u_1^2 \Delta u_2 \Delta P_1 + \Delta u_1^3 \Delta P_2 \right) \right>
\end{equation}
and
\begin{equation}
\left< \Delta u_2 T_{1,3} \right> = \left< \Delta u_2 \left( \Delta u_1 \Delta u_2^2 \Delta P_2 + 3 \Delta u_2^3 \Delta P_1 \right) \right>  .
\end{equation}
$\left< \Delta u_1 T_{4,0} \right>$ is given by multiplying eq.~\eqref{eq_28} by $\Delta u_1$ and averaging, i.e.
\begin{equation}
\left< \Delta u_1 T_{4,0} \right> = 4 \Delta u_1^3 \Delta P_1
\end{equation}
and similarly for $\left< \Delta u_1 T_{2,2} \right>$ and $\left< \Delta u_1 T_{0,4} \right>$. The Laplacian however has the same form as given by Hill, so the coefficients can be carried over. By setting the indices according to the definitions in eq.~\eqref{eq_28}, eq.~\eqref{eq_30} and eq.~\eqref{eq_32} we obtain after averaging the following equations for $\left< T_{4,0} \right>$, $\left< T_{2,2} \right>$ and $\left< T_{0,4} \right>$ in the isotropic form
\begin{align}
\frac{\partial \left< \Delta u_1 T_{4,0} \right>}{\partial r} &+ \frac{2}{r} \left< \Delta u_1 T_{4,0} \right> - \frac{8}{r} \left< \Delta u_2 T_{3,1} \right> = - 12 \left< \Delta u_1 \Delta u_1 \Delta P_1 \Delta P_1  \right> \nonumber \\ 
&+ 2 \nu \left[ \frac{\partial^2 \left< T_{4,0} \right>}{\partial r^2} + \frac{2}{r} \frac{\partial \left< T_{4,0} \right>}{\partial r} - \frac{8}{r^2} \left< T_{4,0} \right> + \frac{24}{r^2} \left< T_{2,2} \right> \right] \nonumber \\
&+ 4 \nu \left< \left(R_1 - R_1' \right) \Delta u_1 \Delta u_1 \Delta u_1 \right> \nonumber \\
&- 4 \left<\Delta P_1 \frac{\partial \Delta u_1 \Delta u_1 \Delta u_1}{\partial t} \right> \nonumber \\
&+ 4 \left< \Delta u_1 \Delta u_1 \Delta u_1 \left( u_n \frac{\partial \Delta P_1}{\partial x_n} + u_n' \frac{\partial \Delta P_1}{\partial x_n'} \right) \right> \\
& - 8 \nu \left< \left( \frac{\partial \Delta u_1 \Delta u_1 \Delta u_1}{\partial x_n} \frac{\partial \Delta P_1}{\partial x_n} + \frac{\partial \Delta u_1 \Delta u_1 \Delta u_1}{\partial x_n'} \frac{\partial \Delta P_1}{\partial x_n'} \right) \right> \nonumber \\
&- 12 \left< \Delta u_1 \Delta P_1 \left( \varepsilon_{11} + \varepsilon_{11}' \right) \right> \nonumber
\end{align}
\begin{align}
\frac{\partial \left< \Delta u_1 T_{2,2} \right>}{\partial r} &+ \frac{2}{r} \left< \Delta u_1 T_{2,2} \right> + \frac{2}{r} \left< \Delta u_2 T_{3,1} \right> - \frac{8}{3 r} \left< \Delta u_2 T_{1,3} \right> = \nonumber \\
&- \left< 2 \Delta u_1 \Delta u_1 \Delta P_2 \Delta P_2 - 8  \Delta u_1 \Delta u_2 \Delta P_1 \Delta P_2 - 2 \Delta u_2 \Delta u_2 \Delta P_1 \Delta P_1 \right> \nonumber \\ 
&+ 2 \nu \left[ \frac{2}{r^2} \left< T_{4,0} \right> + \frac{\partial^2 \left< T_{2,2} \right>}{\partial r^2} + \frac{2}{r} \frac{\partial \left< T_{2,2} \right>}{\partial r} - \frac{14}{r^2} \left< T_{2,2} \right> + \frac{8}{3 r^2} \left< T_{0,4} \right> \right] \nonumber \\
&+ 2 \nu \left< \left(R_1 - R_1' \right) \Delta u_1 \Delta u_2 \Delta u_2 + 2 \nu \left(R_2 - R_2' \right) \Delta u_2 \Delta u_1 \Delta u_1 \right> \nonumber \\
&- 2 \left< \Delta P_1 \frac{\partial \Delta u_1 \Delta u_2 \Delta u_2}{\partial t} - 2 \Delta P_2 \frac{\partial \Delta u_2 \Delta u_1 \Delta u_1}{\partial t} \right> \nonumber \\
&+ 2 \left< \Delta u_1 \Delta u_1 \Delta u_2 \left( u_n \frac{\partial \Delta P_2}{\partial x_n} + u_n' \frac{\partial \Delta P_2}{\partial x_n'} \right) \right> \\
&+ 2 \left< \Delta u_1 \Delta u_2 \Delta u_2 \left( u_n \frac{\partial \Delta P_1}{\partial x_n} + u_n' \frac{\partial \Delta P_1}{\partial x_n'} \right) \right> \nonumber \\
& - 4 \nu \left< \left( \frac{\partial \Delta u_1 \Delta u_1 \Delta u_2}{\partial x_n} \frac{\partial \Delta P_2}{\partial x_n} + \frac{\partial \Delta u_1 \Delta u_1 \Delta u_2}{\partial x_n'} \frac{\partial \Delta P_2}{\partial x_n'} \right. \right. \nonumber \\
&\left. \left. \frac{\partial \Delta u_2 \Delta u_2 \Delta u_1}{\partial x_n} \frac{\partial \Delta P_1}{\partial x_n} + \frac{\partial \Delta u_2 \Delta u_2 \Delta u_1}{\partial x_n'} \frac{\partial \Delta P_1}{\partial x_n'} \right) \right> \nonumber \\
&- \left< 2 \Delta u_1 \Delta P_1 \left( \varepsilon_{22} + \varepsilon_{22}' \right) - 4 \Delta u_1 \Delta P_2 \left( \varepsilon_{12} + \varepsilon_{12}' \right) - 4 \Delta u_2 \Delta P_1 \left( \varepsilon_{12} + \varepsilon_{12}' \right) \right> \nonumber \\
&- \left< 2 \Delta u_2 \Delta P_2 \left( \varepsilon_{11} + \varepsilon_{11}' \right) \right> \nonumber
\end{align}
\begin{align}
\frac{\partial \left< \Delta u_1 T_{0,4} \right>}{\partial r} &+ \frac{2}{r} \left< \Delta u_1 T_{0,4} \right> +\frac{4}{r} \left< \Delta u_2 T_{1,3} \right> = - \left< 12 \Delta u_2 \Delta u_2 \Delta P_2 \Delta P_2 \right> \nonumber \\ 
&+ 2 \nu \left[ \frac{12}{r^2} \left< T_{2,2} \right> + \frac{\partial^2 \left< T_{0,4} \right>}{\partial r^2} + \frac{2}{r} \frac{\partial \left< T_{0,4} \right>}{\partial r} - \frac{4}{r^2} \left< T_{0,4} \right> \right] \nonumber \\
&+ 4 \nu \left< \left(R_2 - R_2' \right) \Delta u_2 \Delta u_2 \Delta u_2 \right> \nonumber \\
&- 4 \left< \Delta P_2 \frac{\partial \Delta u_2 \Delta u_2 \Delta u_2}{\partial t} \right> \\
&+ 4 \left< \Delta u_2 \Delta u_2 \Delta u_2 \left( u_n \frac{\partial \Delta P_2}{\partial x_n} + u_n' \frac{\partial \Delta P_2}{\partial x_n'} \right) \right> \nonumber \\
& - 8 \nu \left< \left( \frac{\partial \Delta u_2 \Delta u_2 \Delta u_2}{\partial x_n} \frac{\partial \Delta P_2}{\partial x_n} + \frac{\partial \Delta u_2 \Delta u_2 \Delta u_2}{\partial x_n'} \frac{\partial \Delta P_2}{\partial x_n'} \right) \right> \nonumber \\
&- 12 \left< \Delta u_2 \Delta P_2 \left( \varepsilon_{22} + \varepsilon_{22}' \right) \right> \nonumber
\end{align}

\section{Fourth order dissipation source terms}

The dissipation source terms in eq.~\eqref{eq_29}, eq.~\eqref{eq_31} and eq.~\eqref{eq_33} have the generic forms  $\Delta u_i \Delta u_j (\varepsilon_{kl} + \varepsilon_{kl}')$. In order to derive the equations for these terms  we first derive an equation for $(\varepsilon_{kl} + \varepsilon_{kl}')$. An equation for the instantaneous value of the product of velocity derivatives is derived by first taking the derivative of the momentum equation written for $u_k$ with respect to $x_m$  and combining it with the corresponding equation for the derivative $\partial u_l / \partial x_m$
\begin{equation}
\begin{aligned}
\frac{\partial}{\partial t} \left( \frac{\partial u_k}{\partial x_m} \right) + u_n \frac{\partial}{\partial x_n} \left( \frac{\partial u_k}{\partial x_m} \right) + \frac{\partial u_n}{\partial x_m}\frac{\partial u_k}{\partial x_n} = - \frac{\partial}{\partial x_m} \left( \frac{\partial p}{\partial x_k} \right) + \nu  \frac{\partial^2}{\partial x_n^2} \left( \frac{\partial u_k}{\partial x_m} \right) ,
\end{aligned}
\end{equation}
resulting in
\begin{equation}
\begin{aligned}
\frac{\partial}{\partial t} \left( \frac{\partial u_k}{\partial x_m} \frac{\partial u_l}{\partial x_m} \right) &+ u_n \frac{\partial}{\partial x_n} \left( \frac{\partial u_k}{\partial x_m} \frac{\partial u_l}{\partial x_m}\right) = - \frac{\partial u_n}{\partial x_m}\frac{\partial u_k}{\partial x_n}\frac{\partial u_l}{\partial x_m} - \frac{\partial u_n}{\partial x_m}\frac{\partial u_k}{\partial x_m}\frac{\partial u_l}{\partial x_n} \\
&- \frac{\partial u_l}{\partial x_m} \frac{\partial^2 p}{\partial x_k \partial x_m} - \frac{\partial u_k}{\partial x_m} \frac{\partial^2 p}{\partial x_l \partial x_m} + \nu  \frac{\partial^2}{\partial x_n^2} \left( \frac{\partial u_k}{\partial x_m} \frac{\partial u_l}{\partial x_m} \right) \\
&- \chi_{kl} ,
\end{aligned}
\end{equation}
where we defined
\begin{equation}
2 \nu \left[\frac{\partial}{\partial x_n} \left( \frac{\partial u_k}{\partial x_m} \right) \frac{\partial}{\partial x_n} \left( \frac{\partial u_l}{\partial x_m} \right) \right] = \chi_{kl} .
\end{equation}
Noticeably, $2 \nu \chi_{ij}$ can be interpreted as the dissipation of $\varepsilon_{ij}$. Using the definitions of  $\varepsilon_{ij}$ and $\varepsilon_{ij}'$ in eq.~\eqref{eq_12} we obtain
\begin{equation}
\begin{aligned}
\frac{\partial}{\partial t} \left( \varepsilon_{kl} + \varepsilon_{kl}' \right) &+ u_n \frac{\partial \varepsilon_{kl} + \varepsilon_{kl}'}{\partial x_n} + u_n' \frac{\partial \varepsilon_{kl} + \varepsilon_{kl}'}{\partial x_n'} = \\
&- 2 \nu \underbrace{\left( \frac{\partial u_n}{\partial x_m}\frac{\partial u_k}{\partial x_n}\frac{\partial u_l}{\partial x_m} + \frac{\partial u_n'}{\partial x_m'}\frac{\partial u_k'}{\partial x_n'}\frac{\partial u_l'}{\partial x_m'} \right)}_{A_{kl} + A_{kl}'} \\
&- 2 \nu \underbrace{\left( \frac{\partial u_n}{\partial x_m}\frac{\partial u_l}{\partial x_n}\frac{\partial u_k}{\partial x_m} + \frac{\partial u_n'}{\partial x_m'}\frac{\partial u_l'}{\partial x_n'}\frac{\partial u_k'}{\partial x_m'} \right)}_{A_{lk} + A_{lk}'} \\
&- 2 \nu \underbrace{\left( \frac{\partial u_k}{\partial x_m}\frac{\partial^2 p}{\partial x_l \partial x_m} + \frac{\partial u_k'}{\partial x_m'}\frac{\partial^2 p'}{\partial x_l' \partial x_m'} \right)}_{P_{kl} + P_{kl}'} \\
&- 2 \nu \underbrace{\left( \frac{\partial u_l}{\partial x_m}\frac{\partial^2 p}{\partial x_k \partial x_m} + \frac{\partial u_l'}{\partial x_m'}\frac{\partial^2 p'}{\partial x_k' \partial x_m'} \right)}_{P_{lk} + P_{lk}'} \\
&+ \nu \left( \frac{\partial^2 \varepsilon_{kl} + \varepsilon_{kl}'}{\partial x_n^2} + \frac{\partial^2 \varepsilon_{kl} + \varepsilon_{kl}'}{\partial x_n'^2} \right) \\
&- 2 \nu \left( \chi_{kl} + \chi_{kl}' \right)
\label{eq_42}
\end{aligned}
\end{equation}
which also defines the quantities $A_{kl} + A_{kl}'$ and $P_{kl} + P_{kl}'$. Combining this with the equation~\eqref{eq_12} one obtains 
\begin{equation}
\begin{aligned}
\frac{\partial \Delta u_i \Delta u_j \left( \varepsilon_{kl} + \varepsilon_{kl}' \right)}{\partial t} &+ u_n \frac{\partial \Delta u_i \Delta u_j \left( \varepsilon_{kl} + \varepsilon_{kl}' \right)}{\partial x_n} + u_n' \frac{\partial \Delta u_i \Delta u_j \left( \varepsilon_{kl} + \varepsilon_{kl}' \right)}{\partial x_n'} = \\
&- 2 \nu \Delta u_i \Delta u_j \left(  A_{kl} + A_{kl}' + A_{lk} + A_{lk}'\right) \\
&- 2 \nu \Delta u_i \Delta u_j \left(  P_{kl} + P_{kl}' + P_{lk} + P_{lk}'\right) \\
&+ \nu \left( \frac{\partial^2 \Delta u_i \Delta u_j \left( \varepsilon_{kl} + \varepsilon_{kl}' \right)}{\partial x_n^2} + \frac{\partial^2 \Delta u_i \Delta u_j \left( \varepsilon_{kl} + \varepsilon_{kl}' \right)}{\partial x_n'^2} \right) \\
&- 2 \nu \left( \frac{\partial \Delta u_i \Delta u_j}{\partial x_n}\frac{\partial \varepsilon_{kl} + \varepsilon_{kl}'}{\partial x_n} + \frac{\partial \Delta u_i \Delta u_j}{\partial x_n'}\frac{\partial \varepsilon_{kl} + \varepsilon_{kl}'}{\partial x_n'} \right) \\
&- \left( \Delta u_j \Delta P_i + \Delta u_i \Delta P_j \right)\left( \varepsilon_{kl} + \varepsilon_{kl}' \right) \\
&- 2 \nu \Delta u_i \Delta u_j \left( \chi_{kl} + \chi_{kl}' \right) \\
&- \left( \varepsilon_{ij} + \varepsilon_{ij}' \right) \left( \varepsilon_{kl} + \varepsilon_{kl}' \right)
\label{eq_43}
\end{aligned}
\end{equation}
The term ($\varepsilon_{ij} + \varepsilon_{ij}') (\varepsilon_{kl} + \varepsilon_{kl}')$ in this equation will, after averaging and using the definitions in eq.~\eqref{eq_29}, eq.~\eqref{eq_31} and eq.~\eqref{eq_33} generate the new dissipation parameters at the fourth order $\left< \varepsilon_{11}^2 \right>$,  $\left< \varepsilon_{12}^2 \right>$, $\left< \varepsilon_{11} \varepsilon_{22} \right>$  and $\left< \varepsilon_{22}^2 \right>$ discussed in the main text. The isotropic form of the transport terms in the equations for $\left< E_{4,0} \right>$, $\left< E_{2,2} \right>$ and $\left< E_{0,4} \right>$ are derived in Appendix~B using the method outlined by Robertson~\cite{robertson1940invariant}. They are
\begin{align}
\frac{\partial \left< \Delta u_1 E_{4,0} \right>}{\partial r} &+ \frac{2}{r} \left< \Delta u_1 E_{4,0} \right> - \frac{8}{r} \left< \Delta u_2 E_{3,1} \right> = \nonumber \\
& -\left< F_{4,0} \right> \nonumber \\
&- 24 \nu  \left< \Delta u_1 \Delta u_1 \left(  P_{11} + P_{11}'\right) \right> \nonumber \\
&+ 2 \nu \left[ \frac{\partial^2 \left< E_{4,0} \right>}{\partial r^2}+ \frac{2}{r} \frac{\partial \left< E_{4,0} \right>}{\partial r} - \frac{8}{r^2} \left< E_{4,0} \right> + \frac{24}{r^2} \left< E_{2,2} \right> \right] \nonumber \\
&- 12 \nu \left< \left( \frac{\partial \Delta u_1 \Delta u_1}{\partial x_n}\frac{\partial \varepsilon_{11}}{\partial x_n} + \frac{\partial \Delta u_1 \Delta u_1}{\partial x_n'}\frac{\varepsilon_{11}'}{\partial x_n'} \right) \right> \label{eq_44}\\
&- 12 \left< \Delta u_1 \Delta P_1 \left( \varepsilon_{11} + \varepsilon_{11}' \right) \right> \nonumber \\
&- 12 \nu \left< \Delta u_1 \Delta u_1 \left( \chi_{11} + \chi_{11}' \right) \right> \nonumber \\
&- 6 \left< \left( \varepsilon_{11} + \varepsilon_{11}' \right)^2 \right> \nonumber
\end{align}
\begin{equation}
E_{31} = 3 \Delta u_2 \Delta u_1 \left( \varepsilon_{11} + \varepsilon_{11}' \right) + 3 \Delta u_1^2 \left( \varepsilon_{12} + \varepsilon_{12}' \right)
\end{equation}
\begin{equation}
F_{4,0} = 24 \nu\Delta u_1 \Delta u_1 \left(  A_{11} + A_{11}'\right)
\end{equation}

\begin{align}
\frac{\partial \left< \Delta u_1 E_{2,2} \right>}{\partial r} &+ \frac{2}{r} \left< \Delta u_1 E_{2,2} \right> + \frac{2}{r} \left< \Delta u_2 E_{3,1} \right> - \frac{8}{3 r} \left< \Delta u_2 E_{1,3} \right> = \nonumber \\
&- \left< F_{2,2} \right> \nonumber \\
&- \left< 2 \nu \left( 2 \Delta u_1 \Delta u_1 \left(  P_{22} + P_{22}' \right) + 4 \Delta u_1 \Delta u_2 \left(  P_{12} + P_{12}' + P_{21} + P_{21}' \right) + 2 \Delta u_2 \Delta u_2 \left(  P_{11} + P_{11}' \right) \right) \right> \nonumber \\
&+ 2 \nu \left[ \frac{2}{r^2} \left< E_{4,0} \right> + \frac{\partial^2 \left< E_{2,2} \right> }{\partial r^2} + \frac{2}{r} \frac{\left< \partial E_{2,2} \right> }{\partial r} - \frac{14}{r^2} \left< E_{2,2} \right> + \frac{8}{3 r^2} \left< E_{0,4} \right> \right] \nonumber \\
&- 2 \nu  \left< \left( \frac{\partial \Delta u_1 \Delta u_1}{\partial x_n}\frac{\partial \varepsilon_{22}}{\partial x_n} + 4 \frac{\partial \Delta u_1 \Delta u_2}{\partial x_n}\frac{\partial \varepsilon_{12}}{\partial x_n} + \frac{\partial \Delta u_2 \Delta u_2}{\partial x_n}\frac{\partial \varepsilon_{11}}{\partial x_n} \right. \right.\\
&+ \left. \left. \frac{\partial \Delta u_1 \Delta u_1}{\partial x_n'}\frac{\partial \varepsilon_{22}'}{\partial x_n'} + 4 \frac{\partial \Delta u_1 \Delta u_2}{\partial x_n'}\frac{\partial \varepsilon_{12}'}{\partial x_n'} + \frac{\partial \Delta u_2 \Delta u_2}{\partial x_n'}\frac{\partial \varepsilon_{11}'}{\partial x_n'} \right) \right> \nonumber \\
&- \left< 2 \Delta u_1 \Delta P_1 \left( \varepsilon_{22} + \varepsilon_{22}' \right) - 4 \left( \Delta u_1 \Delta P_2 + \Delta u_2 \Delta P_1 \right)\left( \varepsilon_{12} + \varepsilon_{12}' \right) - 2 \Delta u_2 \Delta P_2 \left( \varepsilon_{11} + \varepsilon_{11}' \right) \right> \nonumber \\
&- 2 \nu  \left< \Delta u_1 \Delta u_1 \left( \chi_{22} + \chi_{22}' \right) + 4 \Delta u_1 \Delta u_2 \left( \chi_{12} + \chi_{12}' \right) + \Delta u_2 \Delta u_2 \left( \chi_{11} + \chi_{11}' \right)  \right> \nonumber \\
&- \left< 2 \left( \varepsilon_{11} + \varepsilon_{11}' \right) \left( \varepsilon_{22} + \varepsilon_{22}' \right) - 4 \left( \varepsilon_{12} + \varepsilon_{12}' \right) \left( \varepsilon_{12} + \varepsilon_{12}' \right) \right> \nonumber
\end{align}
\begin{equation}
E_{1,3} = 3 \Delta u_1 \Delta u_2 \left( \varepsilon_{22} + \varepsilon_{22}' \right) + 3 \Delta u_2^2 \left( \varepsilon_{12} + \varepsilon_{12}' \right)
\end{equation}
\begin{align}
F_{2,2} =& 2 \nu \left( 2 \Delta u_1 \Delta u_1 \left(  A_{22} + A_{22}'\right) + 4 \Delta u_1 \Delta u_2 \left(  A_{12} + A_{12}' + A_{21} + A_{21}'\right) \right. \\
& \left.+ 2 \Delta u_2 \Delta u_2 \left(  A_{11} + A_{11}'\right) \right) \nonumber
\end{align}
\begin{align}
\frac{\partial \left< \Delta u_1 E_{0,4} \right>}{\partial r} &+ \frac{2}{r} \left< \Delta u_1 E_{0,4} \right> + \frac{4}{r} \left< \Delta u_2 E_{1,3} \right> = \nonumber \\
&- \left< F_{0,4} \right> \nonumber \\
&- 24 \nu \left< \Delta u_2 \Delta u_2 \left(  P_{22} + P_{22}'\right) \right> \nonumber \\
&+ 2 \nu \left[ \frac{12}{r^2} \left< E_{2,2} \right> + \frac{\partial^2 \left< E_{0,4} \right>}{\partial r^2} + \frac{2}{r} \frac{\partial \left< E_{0,4} \right>}{\partial r} - \frac{4}{r^2} \left< E_{0,4} \right> \right] \\
&- 12 \nu \left<  \left( \frac{\partial \Delta u_2 \Delta u_2}{\partial x_n}\frac{\partial \varepsilon_{22}}{\partial x_n} + \frac{\partial \Delta u_2 \Delta u_2}{\partial x_n'}\frac{\partial \varepsilon_{22}'}{\partial x_n'} \right) \right> \nonumber \\
&- 12 \left< \Delta u_2 \Delta P_2 \left( \varepsilon_{22} + \varepsilon_{22}' \right) \right> \nonumber \\
&- 12 \left< \nu \Delta u_2 \Delta u_2 \left( \chi_{22} + \chi_{22}' \right) \right> \nonumber \\
&- 6 \left< \left( \varepsilon_{22} + \varepsilon_{22}' \right)^2 \right> \nonumber
\end{align}
\begin{equation}
F_{0,4} = 24 \nu  \Delta u_2 \Delta u_2 \left(  A_{22} + A_{22}' \right)
\end{equation}

\section{Sixth order}
We also present the equations for the sixth order. The reason for doing this is that the source term of the dissipative source term of the sixth order contains the triple product $\left< \varepsilon_{ij} \varepsilon{kl} \varepsilon_{mn} \right>$, i.e. we find the third moment of the dissipation in the system of equations. In fact, \emph{all} moments of the dissipation are found in the system of equations, when one continues to derive the source terms at higher orders.
 
At the sixth order the structure function read according to Hill (\url{http://arxiv.org/abs/physics/0102055}):
\begin{align}
\left( \frac{\partial}{\partial r} + \frac{2}{r} \right) S_{7,0} &- \frac{12}{r} S_{5,2} = - \left< T_{6,0} \right> - \left< E_{6,0} \right> \nonumber \\
&+ 2 \nu \left[ \frac{\partial^2 S_{6,0}}{\partial r^2}+ \frac{2}{r} \frac{\partial S_{6,0}}{\partial r} - \frac{12}{r^2} S_{6,0} + \frac{60}{r^2} S_{4,2} \right] \\
\frac{\partial S_{4,2}}{\partial t} &+ \left( \frac{\partial}{\partial r} + \frac{4}{r} \right) S_{5,2} - \frac{16}{3 r} S_{3,4} = - \left< T_{4,2} \right> - \left< E_{4,2} \right>  \nonumber \\
&+ 2 \nu \left[ \frac{2}{r^2} S_{6,0} + \frac{\partial^2 S_{4,2}}{\partial r^2} + \frac{2}{r} \frac{\partial S_{4,2}}{\partial r} - \frac{26}{r^2} S_{4,2} + \frac{16}{r^2} S_{2,4} \right] \\
\left( \frac{\partial}{\partial r} + \frac{6}{r} \right) S_{3,4} &- \frac{12}{5 r} S_{1,6}= - \left< T_{2,4} \right> - \left< E_{2,4} \right> \nonumber \\
&+ 2 \nu \left[ \frac{12}{r^2} S_{4,2} + \frac{\partial^2 S_{2,4}}{\partial r^2} + \frac{2}{r} \frac{\partial S_{2,4}}{\partial r} - \frac{24}{r^2} S_{2,4} + \frac{12}{5 r^2} S_{0,6} \right] \\
\left( \frac{\partial}{\partial r} + \frac{8}{r} \right) S_{1,6} &= - \left< T_{0,6} \right> - \left< E_{0,6} \right> \nonumber \\
&+ 2 \nu \left[ \frac{30}{r^2} S_{2,4} + \frac{\partial^2 S_{0,6}}{\partial r^2} + \frac{2}{r} \frac{\partial S_{0,6}}{\partial r} - \frac{6}{r^2} S_{0,6} \right]
\end{align}
\begin{align}
T_{6,0} = & 6 {\Delta u_1}^5 \Delta P_1 \\
E_{6,0} = & 15 {\Delta u_1}^4 \left( \varepsilon_{11} + \varepsilon_{11}' \right) \label{eq_57} \\
T_{4,2} = & 4 {\Delta u_1}^3 {\Delta u_2}^2 \Delta P_1 + 2 {\Delta u_1}^4 \Delta u_2 \Delta P_2 \\
E_{4,2} = & 6 {\Delta u_1}^2 {\Delta u_2}^2 \left( \varepsilon_{11} + \varepsilon_{11}' \right) + 8 {\Delta u_1}^3 \Delta u_2 \left( \varepsilon_{12} + \varepsilon_{12}' \right) + {\Delta u_1}^4 \left( \varepsilon_{22} + \varepsilon_{22}' \right) \label{eq_59} \\
T_{2,4} = & 2 \Delta u_1 {\Delta u_2}^4 \Delta P_1 + 4 {\Delta u_1}^2 {\Delta u_2}^3 \Delta P_2 \\
E_{2,4} = & {\Delta u_2}^4 \left( \varepsilon_{11} + \varepsilon_{11}' \right) + 8 \Delta u_1 {\Delta u_2}^3 \left( \varepsilon_{12} + \varepsilon_{12}' \right) + 6 {\Delta u_1 }^2 {\Delta u_2}^2 \left( \varepsilon_{22} + \varepsilon_{22}' \right) \label{eq_61} \\
T_{0,6} = & 6 {\Delta u_2}^5 \Delta P_2 \\
E_{0,6} = & 15 {\Delta u_2}^4 \left( \varepsilon_{22} + \varepsilon_{22}' \right) \label{eq_63}
\end{align}
We will only consider the dissipation source terms because those will generate the third moment of the dissipation distribution at the end. In order to derive equations for these source terms we first derive an generic equation for the fourth product of instantaneous velocity increments to be called
\begin{equation}
\begin{aligned}
\Delta u^4 = \Delta u_i \Delta u_j \Delta u_k \Delta u_l
\end{aligned}
\end{equation}
Combining eq.~(A.12) for $\Delta u_i \Delta u_j$ with its form for $\Delta u_k \Delta u_l$  one obtains 
\begin{equation}
\begin{aligned}
\frac{\partial {\Delta u}^4}{\partial t} &+ u_n \frac{\partial {\Delta u}^4}{\partial x_n} + u_n' \frac{\partial {\Delta u}^4}{\partial x_n'} = - \Delta u_j \Delta u_k \Delta u_l \Delta P_i - \Delta u_i \Delta u_k \Delta u_l \Delta P_j \\
&- \Delta u_i \Delta u_j \Delta u_l \Delta P_k - \Delta u_i \Delta u_j \Delta u_k \Delta P_l \\
&+ \nu \left( \frac{\partial^2 {\Delta u}^4}{\partial x_n^2} + \frac{\partial^2 {\Delta u}^4}{\partial x_n'^2} \right) \\
&- 2 \nu \left(\frac{\partial \Delta u_i \Delta u_j}{\partial x_n} \frac{\partial \Delta u_k \Delta u_l}{\partial x_n} + \frac{\partial \Delta u_i \Delta u_j}{\partial x_n'} \frac{\partial \Delta u_k \Delta u_l}{\partial x_n'} \right)  \\
&- \Delta u_k \Delta u_l {\left( \varepsilon_{ij} + \varepsilon_{ij}' \right)} \\
&- \Delta u_i \Delta u_j {\left( \varepsilon_{kl} + \varepsilon_{kl}' \right)}
\end{aligned}
\end{equation}
This will be combined with eq.~\eqref{eq_42} to obtain for the combination of terms in eq.~\eqref{eq_57}, eq.~\eqref{eq_59}, eq.~\eqref{eq_61} and eq.~\eqref{eq_63} for the generic form
\begin{equation}
\begin{aligned}
\Delta u^4 \left( \varepsilon_{pq} + \varepsilon_{pq}' \right) = \Delta u_i \Delta u_j \Delta u_k \Delta u_l \left( \varepsilon_{pq} + \varepsilon_{pq}' \right)
\end{aligned}
\end{equation}
the equation
\begin{equation}
\begin{aligned}
\frac{\partial {\Delta u}^4 \left( \varepsilon_{pq} + \varepsilon_{pq}' \right)}{\partial t} &+ u_n \frac{\partial {\Delta u}^4 \left( \varepsilon_{pq} + \varepsilon_{pq}' \right)}{\partial x_n} + u_n' \frac{\partial {\Delta u}^4 \left( \varepsilon_{pq} + \varepsilon_{pq}' \right)}{\partial x_n'} = \\
&- ( \Delta u_j \Delta u_k \Delta u_l \Delta P_i + \Delta u_i \Delta u_k \Delta u_l \Delta P_j + \Delta u_i \Delta u_j \Delta u_l \Delta P_k \\
&+ \Delta u_i \Delta u_j \Delta u_k \Delta P_l )\left( \varepsilon_{pq} + \varepsilon_{pq}' \right) \\
&- 2 \nu {\Delta u}^4 \left(  A_{pq} + A_{pq}' + A_{qp} + A_{qp}'\right) \\
&- 2 \nu {\Delta u}^4 \left(  P_{pq} + P_{pq}' + P_{qp} + P_{qp}'\right) \\
&+ \nu \left( \frac{\partial^2 {\Delta u}^4 \left( \varepsilon_{pq} + \varepsilon_{pq}' \right)}{\partial x_n^2} + \frac{\partial^2 {\Delta u}^4 \left( \varepsilon_{pq} + \varepsilon_{pq}' \right)}{\partial x_n'^2} \right) \\
&- 2 \nu \left( \frac{\partial {\Delta u}^4}{\partial x_n}\frac{\partial \varepsilon_{pq} + \varepsilon_{pq}'}{\partial x_n} + \frac{\partial {\Delta u}^4}{\partial x_n'}\frac{\partial \varepsilon_{pq} + \varepsilon_{pq}'}{\partial x_n'} \right) \\
&- 2 \nu {\Delta u}^4 \left( \chi_{pq} + \chi_{pq}' \right) \\
&- 2 \nu \left( \varepsilon_{pq} + \varepsilon_{pq}' \right) \left( \frac{\partial \Delta u_i \Delta u_j}{\partial x_n}\frac{\partial \Delta u_k \Delta u_l}{\partial x_n} + \frac{\partial \Delta u_i \Delta u_j}{\partial x_n'}\frac{\partial \Delta u_k \Delta u_l}{\partial x_n'} \right) \\
&- \Delta u_k \Delta u_l \left( \varepsilon_{ij} + \varepsilon_{ij}' \right) \left( \varepsilon_{pq} + \varepsilon_{pq}' \right) \\
&- \Delta u_i \Delta u_j \left( \varepsilon_{kl} + \varepsilon_{kl}' \right) \left( \varepsilon_{pq} + \varepsilon_{pq}' \right)
\label{eq_67}
\end{aligned}
\end{equation}
From this form one could derive equations for $\left<E_{6,0}\right>$, $\left<E_{4,2}\right>$, $\left<E_{2,4} \right>$ and $\left<E_{0,6} \right>$. We will not write  those equations down but want to consider a generic equation for the last term on the right hand side of eq.~\eqref{eq_67} to be called 
\begin{equation}
F_{ijklpq} = 15 \Delta u_i \Delta u_j (\varepsilon_{kl} + \varepsilon_{kl}') (\varepsilon_{pq} + \varepsilon_{pq}') .
\label{eq_F}
\end{equation}
For this purpose we first derive an equation for $\varepsilon^2 = (\varepsilon_{kl} + \varepsilon_{kl}') (\varepsilon_{pq} + \varepsilon_{pq}')$ from eq.~\eqref{eq_42} which reads
\begin{equation}
\begin{aligned}
\frac{\partial \varepsilon^2}{\partial t} &+ u_n \frac{\partial \varepsilon^2}{\partial x_n} + u_n' \frac{\partial \varepsilon^2}{\partial x_n'} = \\
&- 2 \nu \left(\varepsilon_{pq} + \varepsilon_{pq}' \right) \left(A_{kl} + A_{kl}' + A_{lk} + A_{lk}' \right) \\
&- 2 \nu \left(\varepsilon_{pq} + \varepsilon_{pq}' \right) \left(P_{kl} + P_{kl}' + P_{lk} + P_{lk}' \right) \\
&- 2 \nu \left(\varepsilon_{kl} + \varepsilon_{kl}' \right) \left(A_{pq} + A_{pq}' + A_{qp} + A_{qp}' \right) \\
&- 2 \nu \left(\varepsilon_{kl} + \varepsilon_{kl}' \right) \left(P_{pq} + P_{pq}' + P_{qp} + P_{qp}' \right) \\
&+ \nu \left( \frac{\partial \varepsilon^2}{\partial x_n^2} + \frac{\partial \varepsilon^2}{\partial x_n'^2} \right) \\
&- 2 \nu \left( \frac{\partial \varepsilon_{kl} + \varepsilon_{kl}'}{\partial x_n} \frac{\partial \varepsilon_{pq} + \varepsilon_{pq}'}{\partial x_n} + \frac{\partial \varepsilon_{kl} + \varepsilon_{kl}'}{\partial x_n'} \frac{\partial \varepsilon_{pq} + \varepsilon_{pq}'}{\partial x_n'} \right) \\
&- 2 \nu \left( \varepsilon_{kl} + \varepsilon_{kl}' \right) \left( \chi_{pq} + \chi_{pq}' \right) \\
&- 2 \nu \left( \varepsilon_{pq} + \varepsilon_{pq}' \right) \left( \chi_{kl} + \chi_{kl}' \right)
\end{aligned}
\end{equation}
Combining this equation with eq.~\eqref{eq_12} one obtains after averaging for $\left<\Delta u_i \Delta u_j \varepsilon^2 \right>=$

$\left<\Delta u_i \Delta u_j (\varepsilon_{kl} + \varepsilon_{kl}') (\varepsilon_{pq} + \varepsilon_{pq}') \right>$ the form
\begin{equation}
\begin{aligned}
\left< u_n \frac{\partial \Delta u_i \Delta u_j \varepsilon^2}{\partial x_n} \right. &+ \left. u_n' \frac{\partial \Delta u_i \Delta u_j \varepsilon^2}{\partial x_n'} \right> = \\
&- \left< \varepsilon^2 \left( \Delta u_j \Delta P_i + \Delta u_i \Delta P_j \right) \right> \\
&- 2 \nu \left< \Delta u_i \Delta u_j \left(\varepsilon_{pq} + \varepsilon_{pq}' \right) \left(A_{kl} + A_{kl}' + A_{lk} + A_{lk}' \right) \right> \\
&- 2 \nu \left< \Delta u_i \Delta u_j \left(\varepsilon_{pq} + \varepsilon_{pq}' \right) \left(P_{kl} + P_{kl}' + P_{lk} + P_{lk}' \right) \right> \\
&- 2 \nu \left< \Delta u_i \Delta u_j \left(\varepsilon_{kl} + \varepsilon_{kl}' \right) \left(A_{pq} + A_{pq}' + A_{qp} + A_{qp}' \right) \right>\\
&- 2 \nu \left< \Delta u_i \Delta u_j \left(\varepsilon_{kl} + \varepsilon_{kl}' \right) \left(P_{pq} + P_{pq}' + P_{qp} + P_{qp}' \right) \right> \\
&+ \nu \left< \left( \frac{\partial^2 \Delta u_i \Delta u_j \varepsilon^2}{\partial x_n^2} + \frac{\partial^2 \Delta u_i \Delta u_j \varepsilon^2}{\partial x_n'^2} \right) \right> \\
&- 2 \nu \left< \left( \frac{\partial \Delta u_i \Delta u_j}{\partial x_n} \frac{\partial \varepsilon^2}{\partial x_n} + \frac{\partial \Delta u_i \Delta u_j}{\partial x_n'} \frac{\partial \varepsilon^2}{\partial x_n'} \right) \right> \\
&- 2 \nu \left< \Delta u_i \Delta u_j \left( \frac{\partial \varepsilon_{kl}}{\partial x_n} \frac{\partial \varepsilon_{pq}}{\partial x_n} + \frac{\partial \varepsilon_{kl}'}{\partial x_n'} \frac{\partial \varepsilon_{pq}'}{\partial x_n'} \right) \right> \\
&- 2 \nu  \left< \Delta u_i \Delta u_j \left( \varepsilon_{kl} + \varepsilon_{kl}' \right) \left( \chi_{pq} + \chi_{pq}' \right) \right> \\
&- 2 \nu \left< \Delta u_i \Delta u_j \left( \varepsilon_{pq} + \varepsilon_{pq}' \right) \left( \chi_{kl} + \chi_{kl}' \right) \right> \\
&- \left< \left( \varepsilon_{ij} + \varepsilon_{ij}' \right) \left( \varepsilon_{kl} + \varepsilon_{kl}' \right) \left( \varepsilon_{pq} + \varepsilon_{pq}' \right) \right>
\label{eq_69}
\end{aligned}
\end{equation}
In this equation the term $\left< (\varepsilon_{ij} + \varepsilon_{ij}') (\varepsilon_{kl} + \varepsilon_{kl}') (\varepsilon_{pq} + \varepsilon_{pq}') \right>$ will generate third moments of the dissipation distribution. The longitudinal form $\left< F_{6,0} \right> = 15 \left< \Delta u_1^2 (\varepsilon_{11} + \varepsilon_{11}')^2 \right> $ of eq.~\eqref{eq_69} will contain the term $15 \left<(\varepsilon_{11} + \varepsilon_{11}')^3 \right>$ which generate  parameter $\left< \varepsilon_{11}^3 \right>$ for large $r$.

\section{Discussion}
The procedure of deriving equations which parameters proportional to $\left< \varepsilon^n \right>$ is based on identifying source terms where derivatives square are multiplied by $\Delta u_i \Delta u_j$. A first example is the fourth order dissipation source term $E_{4,0} = 6 \Delta u_1^2 (\varepsilon_{11} + \varepsilon_{11}')$ in eq.~\eqref{eq_29} which led to the term $6 \left<(\varepsilon_{11} + \varepsilon_{11}')^2\right>$ in eq.~\eqref{eq_44} and finally to the dissipation parameter $\left<\varepsilon_{11}^2\right>$. Other  source terms which could be considered are  $\Delta u_i \Delta u_j \Delta P_k \Delta P_l$ in eq.~\eqref{eq_36} or the terms $ 2 \nu \Delta u_i \Delta u_j (A_{kl} + A_{kl}')$ and $2 \nu \Delta u_i \Delta u_j (\chi_{kl} + \chi_{kl}')$ in eq.~\eqref{eq_43}. In the equation for $\Delta u_i \Delta u_j \Delta P_k \Delta P_l$ parameters $(\varepsilon_{ij} + \varepsilon_{ij}') \Delta P_k \Delta P_l$ then appear which are proportional to velocity derivatives squared times pressure gradients squared. In the equations for the F-terms (eq.~\eqref{eq_F}) we find terms which are the product of five velocity derivatives. Those  become constant in the inertial range but not in the dissipative range. Their contribution to intermittency are not clear at this stage.

%\subsection*{References}

%\begin{list}{[\arabic{qcounter}]}{\usecounter{qcounter}}
%\item R.J. Hill, Exact second order structure function relationships, J.  Fluid Mech. 468, 317-326 (2002)
%\item R.J. Hill, Equations relating structure functions of all orders, J. Fluid Mech. 434, 379-388 (2001)
%\item H. Robertson, The invariant theory of isotropic turbulence, Math. Proc.  Cambridge Phil. Soc. 36, 209-223 (1940)
%\item A. N. Kolmogorov, Dokl. Akad. Nauk. SSSR 32, 16 (1941)

%\end{list}

\begin{acknowledgments}
This work was supported by the Deutsche Forschungsgemeinschaft through the grant Pe 241/30-3. The authors gratefully acknowledges the computing time granted by the JARA-HPC-Vergabegremium provided on the JARA-HPC Partition part of the supercomputer JUQUEEN at the Forschungszentrum J\"ulich.
\end{acknowledgments}

% Create the reference section using BibTeX:
\bibliography{arxiv_A.bib}

\appendix
\section{Laplacian of fourth order structure functions}
Following Robertson~\cite{robertson1940invariant}, a fourth order tensor of two-point type which is invariant to rotation and reflection of the coordinate system (cf. isotropic turbulence) is given by

\begin{equation}
\begin{aligned}
A_{ijkl} &= A_1 \frac{r_i r_j r_k r_l}{r^4} + A_2 \delta_{ij} \frac{r_k r_l}{r^2} + A_3 \delta_{ik} \frac{r_j r_l}{r^2} + A_4 \delta_{il} \frac{r_j r_k}{r^2} + A_5 \delta_{jk} \frac{r_i r_l}{r^2} + A_6 \delta_{jl} \frac{r_i r_k}{r^2} + A_7 \delta_{kl} \frac{r_i r_j}{r^2} \\
&+ A_8 \delta_{ij} \delta_{kl} + A_9 \delta_{ik} \delta_{jl} + A_{10} \delta_{jk} \delta_{il} ,
\label{general_form_tensor_4_order}
\end{aligned}
\end{equation}
where $A_i$ are scalar functions, $\delta_{ij}$ is the Kronecker delta, i.e. $\delta_{ij} = 1$ for $i=j$ and $\delta_{ij} = 0$ for $i \neq j$ and $r_i$ a separation vector with magnitude $r$, i.e. $x_i = x_i' + r_i$ where $x_i$ and $x_i'$ are two points in space.

In the following, let $A_{ijkl} = S_{ijkl} = \left< \Delta u_i \Delta u_j \Delta u_k \Delta u_l \right>$, where $\Delta u = u_i - u_i'$, i.e the fourth order structure function. As $S_{ijkl} = S_{jikl} = S_{ijlk} = S_{jlik} =  ...$, $A_2 = A_3 = ... = A_7 = S_2$ and $A_8 = A_9 = A_{10} = S_3$, i.e. from eq.~\eqref{general_form_tensor_4_order}
\begin{equation}
\begin{aligned}
S_{ijkl} &= S_1 \frac{r_i r_j r_k r_l}{r^4} + S_2 \left(\delta_{ij} \frac{r_k r_l}{r^2} + \delta_{ik} \frac{r_j r_l}{r^2} + \delta_{il} \frac{r_j r_k}{r^2} + \delta_{jk} \frac{r_i r_l}{r^2} + \delta_{jl} \frac{r_i r_k}{r^2} + \delta_{kl} \frac{r_i r_j}{r^2} \right) \\
&+ S_3 \left(\delta_{ij} \delta_{kl} + \delta_{ik} \delta_{jl} + \delta_{jk} \delta_{il} \right). 
\label{SF_tensor_4_order}
\end{aligned}
\end{equation}

Next, the scalar functions $S_1$, $S_2$, $S_3$ need to be determined. Without loss of generality, let $r_1 = r$, $r_2 = r_3 = 0$. Choosing $S_{1111}$, eq.~\eqref{SF_tensor_4_order} then yields
\begin{equation}
\begin{aligned}
S_{1111} = S_{4,0} = S_1 + 6 S_2 + 3 S_3 ,
\end{aligned}
\end{equation}
while $S_{1122}$ gives
\begin{equation}
\begin{aligned}
S_{1122} = S_{2,2} = S_2 + S_3
\end{aligned}
\end{equation}
and $S_{2222}$
\begin{equation}
\begin{aligned}
S_{2222} = S_{0,4} = 3 S_3 .
\end{aligned}
\end{equation}
Thus, the three scalar functions $S_1$, $S_2$ and $S_3$ are determined by the three tensor components $S_{1111}$, $S_{1122}$ and $S_{2222}$ and solving for them gives
\begin{equation}
\begin{aligned}
S_1 &= S_{1111} - 6 S_{1122} + S_{2222} , \\
S_2 &= S_{1122} - \frac{1}{3} S_{2222}, \\
S_3 &= \frac{1}{3} S_{2222} .
\label{rel_S_i}
\end{aligned}
\end{equation}

Next, $\partial^2 S_{ijkl} / \partial r_n^2$ depending on the scalar functions $S_1$, $S_2$, $S_3$ and the separation vector $r_i$ is derived, for which the following relations are needed
\begin{equation}
\begin{aligned}
\frac{r_n r_n}{r^2} &= 1 \\
\frac{\partial}{\partial r_n} \left( \frac{r_i}{r} \right) &= \left(\delta_{in} - \frac{r_i r_n}{r^2} \right) \frac{1}{r} \\
\frac{\partial}{\partial r_n} \left( \frac{r_n}{r} \right) &= \left(\delta_{nn} - \frac{r_n r_n}{r^2} \right) \frac{1}{r} = \frac{2}{r} \\
\frac{\partial A(r)}{\partial r_n} &= \frac{r_n}{r} \frac{\partial A(r)}{\partial r} \\
\frac{\partial}{\partial r_n} \left( \frac{1}{r} \right) &= -\frac{1}{r^2} \frac{\partial \sqrt{r_i^2}}{\partial r_n} = -\frac{1}{r^2} \frac{r_i}{\sqrt{r_i^2}} \frac{\partial r_i}{\partial r_n} = -\frac{1}{r^2} \frac{r_i}{r} \delta_{in} = - \frac{r_n}{r^3}
\label{simpl_1}
\end{aligned}
\end{equation}
Consequently,
\begin{equation}
\begin{aligned}
\frac{r_n}{r} \frac{\partial}{\partial r_n} \left( \frac{r_i}{r} \right) = \left(r_n \delta_{in} - \frac{r_n r_i r_n}{r^2} \right) \frac{1}{r^2} = \left(r_i - r_i \right) \frac{1}{r^2} = 0 .
\label{simpl_2}
\end{aligned}
\end{equation}

Using eq.~\eqref{simpl_1} and eq.~\eqref{simpl_2} then results in
\begin{equation}
\begin{aligned}
\frac{\partial^2}{\partial r_n^2} \left( S_1 \frac{r_i r_j r_k r_l}{r^4} \right) &= \frac{r_i r_j r_k r_l}{r^4} \left( \frac{\partial^2 S_1}{\partial r^2} + \frac{2}{r} \frac{\partial S_1}{\partial r} \right) \\
&+ \frac{2}{r^2} S_1  \left[ \frac{r_k r_l}{r^2} \left( \delta_{ij} - \frac{r_i r_j}{r^2} \right) + \frac{r_j r_l}{r^2} \left( \delta_{ik} - \frac{r_i r_k}{r^2} \right) + \frac{r_j r_k}{r^2} \left( \delta_{il} - \frac{r_i r_l}{r^2} \right) \right. \\
&+ \left. \frac{r_i r_l}{r^2} \left( \delta_{jk} - \frac{r_j r_k}{r^2} \right) + \frac{r_i r_k}{r^2} \left( \delta_{jl} - \frac{r_j r_l}{r^2} \right) + \frac{r_i r_j}{r^2} \left( \delta_{kl} - \frac{r_k r_l}{r^2} \right) \right] \\
&- \frac{8}{r^2} S_1 \frac{r_i r_j r_k r_l}{r^4} .
\label{laplace_D1}
\end{aligned}
\end{equation}

Similarly with eq.~\eqref{simpl_1} and eq.~\eqref{simpl_2}
\begin{equation}
\begin{aligned} 
\frac{\partial^2}{\partial r_n^2} \left[ S_2 \left(\delta_{ij} \frac{r_k r_l}{r^2}  \right. \right. &+ \left. \left. \delta_{ik} \frac{r_j r_l}{r^2} + \delta_{il} \frac{r_j r_k}{r^2} + \delta_{jk} \frac{r_i r_l}{r^2} + \delta_{jl} \frac{r_i r_k}{r^2} + \delta_{kl} \frac{r_i r_j}{r^2} \right) \right] \\
&= \left(\delta_{ij} \frac{r_k r_l}{r^2} + \delta_{ik} \frac{r_j r_l}{r^2} + \delta_{il} \frac{r_j r_k}{r^2} + \delta_{jk} \frac{r_i r_l}{r^2} \right. \\
&+ \left. \delta_{jl} \frac{r_i r_k}{r^2} + \delta_{kl} \frac{r_i r_j}{r^2} \right) \left( \frac{\partial^2 S_2}{\partial r^2} + \frac{2}{r}\frac{\partial S_2}{\partial r}\right) \\
&+ 2 \frac{S_2}{r^2} \left[ \delta_{ij} \left( \left( \delta_{kl} - \frac{r_k r_l}{r^2} \right) - 2 \frac{r_k r_l}{r^2} \right) + \delta_{ik} \left( \left( \delta_{jl} - \frac{r_j r_l}{r^2} \right) - 2 \frac{r_j r_l}{r^2} \right) \right. \\
&+ \left. \delta_{il} \left( \left( \delta_{jk} - \frac{r_j r_k}{r^2} \right) - 2 \frac{r_j r_k}{r^2} \right) + \delta_{jk} \left( \left( \delta_{il} - \frac{r_i r_l}{r^2} \right) - 2 \frac{r_j r_l}{r^2} \right) \right. \\
&+ \left. \delta_{jl} \left( \left( \delta_{ik} - \frac{r_i r_k}{r^2} \right) - 2 \frac{r_i r_k}{r^2} \right) + \delta_{kl} \left( \left( \delta_{ij} - \frac{r_i r_j}{r^2} \right) - 2 \frac{r_i r_j}{r^2} . \right) \right]
\label{laplace_D2}
\end{aligned}
\end{equation}

Finally,
\begin{equation}
\begin{aligned}
\frac{\partial^2}{\partial r_n^2} \left[ S_3 \left(\delta_{ij} \delta_{kl} + \delta_{ik} \delta_{jl} + \delta_{jk} \delta_{il} \right) \right] = \left(\delta_{ij} \delta_{kl} + \delta_{ik} \delta_{jl} + \delta_{jk} \delta_{il} \right) \left( \frac{\partial^2 S_3}{\partial r^2} + \frac{2}{r}\frac{\partial S_3}{\partial r}\right) .
\label{laplace_D3}
\end{aligned}
\end{equation}

Now, adding eq.~\eqref{laplace_D1}, eq.~\eqref{laplace_D2} and eq.~\eqref{laplace_D3} and substituting $i=j=k=l=1$ gives
\begin{equation}
\begin{aligned}
\frac{\partial^2 S_{1111}}{\partial r_n^2} &= \left( \frac{\partial^2 S_1}{\partial r^2} + \frac{2}{r}\frac{\partial S_1}{\partial r}\right) - \frac{8}{r^2}S_1 \\
&+ 6 \left( \frac{\partial^2 S_2}{\partial r^2} + \frac{2}{r}\frac{\partial S_2}{\partial r}\right) - \frac{24}{r^2}S_2 \\
&+ 3 \left( \frac{\partial^2 S_3}{\partial r^2} + \frac{2}{r}\frac{\partial S_3}{\partial r}\right) .
\end{aligned}
\end{equation}
Using the relations eq.~\eqref{rel_S_i} yields
\begin{equation}
\begin{aligned}
\frac{\partial^2 S_{1111}}{\partial r_n^2} &= \left( \frac{\partial^2 S_{1111}}{\partial r^2} + \frac{2}{r}\frac{\partial S_{1111}}{\partial r}\right) - \frac{8}{r^2}S_{1111} + \frac{24}{r^2}S_{1122} .
\label{laplace_S1111}
\end{aligned}
\end{equation}

Setting $i=j=1$, $k=l=2$ in the summation of eq.~\eqref{laplace_D1}, eq.~\eqref{laplace_D2} and eq.~\eqref{laplace_D3} yields 
\begin{equation}
\begin{aligned}
\frac{\partial^2 S_{1122}}{\partial r_n^2} &= \frac{2}{r^2}S_1 + \left( \frac{\partial^2 S_2}{\partial r^2} + \frac{2}{r}\frac{\partial S_2}{\partial r}\right) - \frac{2}{r^2}S_2 \\
&+ \left( \frac{\partial^2 S_3}{\partial r^2} + \frac{2}{r}\frac{\partial S_3}{\partial r}\right)
\end{aligned}
\end{equation}
and with the relations eq.~\eqref{rel_S_i}
\begin{equation}
\begin{aligned}
\frac{\partial^2 S_{1122}}{\partial r_n^2} &= \left( \frac{\partial^2 S_{1122}}{\partial r^2} + \frac{2}{r}\frac{\partial S_{1122}}{\partial r}\right) + \frac{2}{r^2}S_{1111} - \frac{14}{r^2}S_{1122} + \frac{8}{3 r^2}S_{2222} .
\label{laplace_S1122}
\end{aligned}
\end{equation}

In the same way, setting $i=j=k=l=2$ gives
\begin{equation}
\begin{aligned}
\frac{\partial^2 S_{2222}}{\partial r_n^2} &= \frac{12}{r^2}S_2 + 3 \left( \frac{\partial^2 S_3}{\partial r^2} + \frac{2}{r}\frac{\partial S_3}{\partial r}\right) ,
\end{aligned}
\end{equation}
i.e. using eq.~\eqref{rel_S_i}
\begin{equation}
\begin{aligned}
\frac{\partial^2 S_{2222}}{\partial r_n^2} &= \left( \frac{\partial^2 S_{2222}}{\partial r^2} + \frac{2}{r}\frac{\partial S_{2222}}{\partial r}\right) + \frac{12}{r^2} S_{1122} - \frac{4}{r^2}S_{2222} .
\label{laplace_S2222}
\end{aligned}
\end{equation}

These are the same equations Hill~\cite{hill2002exact} derived using a matrix algorithm (note there is a recent correction to the matrix algorithm, available at http://arxiv.org/abs/physics/0102055).

\section{Divergence of the fourth order dissipation source terms}
Again following Robertson~\cite{robertson1940invariant}, the general expression for a fifth order tensor of two-point type is given by
\begin{equation}
\begin{aligned}
B_{nijkl} &= B_1 \frac{r_n}{r}\frac{r_i}{r}\frac{r_j}{r}\frac{r_k}{r}\frac{r_l}{r} \\
&+ B_2 \delta_{ni}\frac{r_j}{r}\frac{r_k}{r}\frac{r_l}{r} + B_3 \delta_{nj}\frac{r_i}{r}\frac{r_k}{r}\frac{r_l}{r} + B_4 \delta_{nk}\frac{r_i}{r}\frac{r_j}{r}\frac{r_l}{r} + B_5 \delta_{nl}\frac{r_i}{r}\frac{r_j}{r}\frac{r_k}{r} + B_6 \delta_{ij}\frac{r_n}{r}\frac{r_k}{r}\frac{r_l}{r} \\
&+ B_7 \delta_{ik}\frac{r_n}{r}\frac{r_j}{r}\frac{r_l}{r} + B_8 \delta_{il}\frac{r_n}{r}\frac{r_j}{r}\frac{r_k}{r} + B_9 \delta_{jk}\frac{r_n}{r}\frac{r_i}{r}\frac{r_l}{r} + B_{10} \delta_{jl}\frac{r_n}{r}\frac{r_i}{r}\frac{r_k}{r} + B_{11} \delta_{kl}\frac{r_n}{r}\frac{r_i}{r}\frac{r_j}{r} \\
&+ B_{12} \delta_{ni} \delta_{jk} \frac{r_l}{r} + B_{13} \delta_{ni} \delta_{jl} \frac{r_k}{r} + B_{14} \delta_{ni} \delta_{kl} \frac{r_j}{r} + B_{15} \delta_{nj} \delta_{ik} \frac{r_l}{r} + B_{16} \delta_{nj} \delta_{il} \frac{r_k}{r} \\
&+ B_{17} \delta_{nj} \delta_{kl} \frac{r_i}{r} + B_{18} \delta_{nk} \delta_{ij} \frac{r_l}{r} + B_{19} \delta_{nk} \delta_{il} \frac{r_j}{r} + B_{20} \delta_{nk} \delta_{jl} \frac{r_i}{r} + B_{21} \delta_{nl} \delta_{ij} \frac{r_k}{r} \\
&+ B_{22} \delta_{nl} \delta_{ik} \frac{r_j}{r} + B_{23} \delta_{nl} \delta_{jk} \frac{r_i}{r} + B_{24} \delta_{ij} \delta_{kl} \frac{r_n}{r} + B_{25} \delta_{ik} \delta_{jl} \frac{r_n}{r} + B_{26} \delta_{il} \delta_{jk} \frac{r_n}{r} .
\label{eq_E}
\end{aligned}
\end{equation}
We consider tensors of the form $\left<E_{n,ijkl}\right> = \left<\Delta u_n E_{ijkl}\right>$,
\begin{equation}
\begin{aligned}
\left< \Delta u_n E_{ijkl} \right> &= \left< \Delta u_n \left[ \Delta u_i \Delta u_j \left( \varepsilon_{kl} + \varepsilon_{kl'} \right) + \Delta u_i \Delta u_k \left( \varepsilon_{jl} + \varepsilon_{jl'} \right) + \Delta u_i \Delta u_l \left( \varepsilon_{jk} + \varepsilon_{jk'} \right) \right. \right.\\
& \left. \left. + \Delta u_j \Delta u_k \left( \varepsilon_{il} + \varepsilon_{il'} \right) + \Delta u_j \Delta u_l \left( \varepsilon_{ik} + \varepsilon_{ik'} \right) + \Delta u_k \Delta u_l \left( \varepsilon_{ij} + \varepsilon_{ij'} \right) \right] \right> ,
\end{aligned}
\end{equation}
where the indices $i$, $j$, $k$ and $l$ may be interchanged, but $n$ may not. Consequently, $B_1 = E_1$, $B_2 = ... = B_5 = E_2$, $B_6 = ... = B_{11} = E_3$, $B_{12} = ... = B_{23} = E_4$ and $B_{24} = B_{25} = B_{26} = E_5$. Thus we end up with five scalar functions of $r$ instead of three as we may not interchange all indices. That is, we may not use the result of Hill~\cite{hill2002exact} for the divergence of the fourth order structure functions. Next, we need to determine $E_i$. We choose
\begin{equation}
\begin{aligned}
\left< \Delta u_1 E_{4,0} \right> = \left< E_{1,1111} \right> &= E_1 + 4 E_2 + 6 E_3 + 12 E_4 + 3 E_5 \\
\left< \Delta u_1 E_{2,2} \right> = \left< E_{1,1122} \right> &= E_3 + 2 E_4 + E_5 \\
\left< \Delta u_1 E_{0,4} \right> = \left< E_{1,2222} \right> &= 3 E_5 \\
\left< \Delta u_2 E_{3,1} \right> = \left< E_{2,2111} \right> &= E_2 + 3 E_4 \\
\left< \Delta u_2 E_{1,3} \right> = \left< E_{2,1222} \right> &= 3 E_4
\label{eq_E_i}
\end{aligned}
\end{equation}
and solving for the $E_i$ gives
\begin{equation}
\begin{aligned}
E_1 &= \left< E_{1,1111} \right> - 6 \left< E_{1,1122} \right> + \left< E_{1,2222} \right> - 4 \left< E_{2,2111} \right> + 4 \left< E_{2,1222} \right> \\
E_2 &= \left< E_{2,2111} \right> - \left< E_{2,1222} \right> \\
E_3 &= \left< E_{1,1122} \right> - \frac{2}{3} \left< E_{2,1222} \right> - \frac{1}{3} \left< E_{1,2222} \right> \\
E_4 &= \frac{1}{3} \left< E_{2,1222} \right> \\
E_5 &= \frac{1}{3} \left< E_{1,2222} \right> .
\end{aligned}
\end{equation}
Taking the derivative of eq.~\eqref{eq_E} using the relations  eq.~\eqref{simpl_1} then results in 
\begin{equation}
\begin{aligned}
\frac{\partial}{\partial r_n} \left( \left< E_{n,ijkl} \right> \right) &= \left( \frac{\partial E_1}{\partial r} + \frac{2}{r} E_1 + 4 \frac{\partial E_2}{\partial r} - \frac{12}{r} E_2 \right) \frac{r_i r_j r_k r_l}{r^4} \\
&+ \left( \frac{2}{r} E_2 + \frac{\partial E_3}{\partial r} + \frac{2}{r} E_3 + 2 \frac{\partial E_4}{\partial r} - \frac{2}{r} E_4 \right) \left( \delta_{ij} \frac{r_k r_l}{r^2} + \delta_{ik} \frac{r_j r_l}{r^2} + \delta_{il} \frac{r_j r_k}{r^2} \right. \\
&\left. + \delta_{jk} \frac{r_i r_l}{r^2} + \delta_{jl} \frac{r_i r_k}{r^2} + \delta_{kl} \frac{r_i r_j}{r^2} \right) \\
& +\left( \frac{4}{r} E_4 + \frac{\partial E_5}{\partial r} + \frac{2}{r} E_5 \right) \left( \delta_{ij} \delta_{kl} + \delta_{ik} \delta_{jl} + \delta_{il} \delta_{jk} \right) .
\end{aligned}
\end{equation}
Therefore using eq.~\eqref{eq_E_i},
\begin{equation}
\begin{aligned}
\frac{\partial \left< E_{n,1111} \right>}{\partial r} &= \frac{\partial \left< E_{1,1111} \right>}{\partial r} + \frac{2}{r} \left< E_{1,1111} \right> - \frac{8}{r} \left< E_{2,2111} \right> \\
\frac{\partial \left< E_{n,1122} \right>}{\partial r} &= \frac{\partial \left< E_{1,1122} \right>}{\partial r} + \frac{2}{r} \left< E_{1,1122} \right> + \frac{2}{r} \left< E_{2,2111} \right> - \frac{8}{3 r} \left< E_{2,1222} \right> \\
\frac{\partial \left< E_{n,2222} \right>}{\partial r} &= \frac{\partial \left< E_{1,2222} \right>}{\partial r} + \frac{2}{r} \left< E_{1,2222} \right> + \frac{4}{r} \left< E_{2,1222} \right> ,
\end{aligned}
\end{equation}
where
\begin{equation}
\begin{aligned}
\left< E_{1,1111} \right> = \left< \Delta u_1 E_{4,0} \right> = 6 \left< \Delta u_1^3 \left( \varepsilon_{11} + \varepsilon_{11}' \right) \right>
\end{aligned}
\end{equation}
\begin{equation}
\begin{aligned}
\left< E_{1,1122} \right> = \left< \Delta u_1 E_{2,2} \right>  = \left< \Delta u_1 \left[ \Delta u_1^2 \left( \varepsilon_{22} + \varepsilon_{22}' \right) + 4 \Delta u_1 u_2 \left( \varepsilon_{12} + \varepsilon_{12}' \right) + \Delta u_2^2 \left( \varepsilon_{11} + \varepsilon_{11}' \right) \right] \right> 
\end{aligned}
\end{equation}
\begin{equation}
\begin{aligned}
\left< E_{1,2222} \right> = \left< \Delta u_1 E_{0,4} \right> = 6 \left< \Delta u_1 \Delta u_2^2 \left( \varepsilon_{22} + \varepsilon_{22}' \right) \right>
\end{aligned}
\end{equation}
\begin{equation}
\begin{aligned}
\left< E_{2,2111} \right> = \left< \Delta u_2 E_{3,1} \right> = \left< \Delta u_2 \left[ 3 \Delta u_2 \Delta u_1 \left( \varepsilon_{11} + \varepsilon_{11}' \right) + 3 \Delta u_1^2 \left( \varepsilon_{12} + \varepsilon_{12}' \right) \right] \right>
\end{aligned}
\end{equation}
\begin{equation}
\begin{aligned}
\left< E_{2,122} \right> = \left< \Delta u_2 E_{1,3} \right> = \left< \Delta u_2 \left[ 3 \Delta u_1 \Delta u_2 \left( \varepsilon_{22} + \varepsilon_{22}' \right) + 3 \Delta u_2^2 \left( \varepsilon_{12} + \varepsilon_{12}' \right) \right] \right> .
\end{aligned}
\end{equation}

\end{document}